Raja Ramanna

\documentclass[12pt]{article}
\begin{document}

\begin{center}
{\bf  THE SYSTEMATICS OF FUNDAMENTAL PARTICLES  AND UNSTABLE NUCLEAR
SYSTEMS  USING THE CONCEPT OF CONTINUITY AND DISCRETENESS 
}		
	
\vspace*{0.5cm}
{\bf Raja Ramanna and Anju Sharma \\
NATIONAL INSTITUTE OF ADVANCED STUDIES \\
BANGALORE}
\end{center}

\noindent{\large{\bf \underline{ABSTRACT:}}}
\vspace*{0.3cm}

In this review, a  new method is proposed  based on Cantor's theory of
Cardinality to  analyze the  experimental   data on  unstable  nuclear
systems  which includes   fundamental  particles and  their  flavours,
$\beta$-decay including  spin   and parity,  $\alpha$-decay  and   the
systematics of  the energy  levels  of light  nuclei.  The method also
derives a  formula for the Binding Energies  of unstable  nuclei.  All
this is based on the theory of discreteness and continuity. 

\section{ Introduction:} 

The earliest review on nuclear  physics was written nearly sixty years
ago by H.  A.  Bethe and his associates [1] to explain various nuclear
phenomena within the framework   of quantum theory.  The review  which
was  published in a  series of  three   articles which  incorporated a
thorough and  comprehensive study of the  developments in the field of
nuclear physics at that time.  The review  covered almost every aspect
of nucleus ranging  from $\alpha$-, $\beta$-, $\gamma$- radioactivity,
nuclear reactions, nuclear forces and many body effects etc. 

Bethe's  work was   epoch-making  and influenced  all  other  research
subsequent to his  paper.  For many years it  was considered  that the
nuclear  forces  obtained by   various experiments  was  sufficient to
understand all nuclear phenomena   but unfortunately the  strong  many
body nature of the  problem  shifted the  fundamental approach to  the
study  of nuclear   models and has   been the  approach of all   later
workers.  Though enormous  developments  have taken place  in  nuclear
sciences, such  as the discovery of  magic  numbers, excited states of
nuclei up to high energies, discovery of  parity to mention only a few
that fall in  this  category, an overall  picture of  nuclear  systems
which includes fundamental  particles  is  still incomplete and   will
probably remain as such. 

Even though   there  has been a  continuous  progress  in dealing with
nuclear systems and fundamental particles, there still is  a lack of a
coherent  formulation which would permit  an analysis  of all observed
phenomena in a fundamental way.  In  place of a single unifying theory
there are scattered  islands  of   knowledge in  a  sea of   seemingly
uncorrelated observations.  So    far,  attempts have   been made   to
formulate several theories  and models to  understand various physical
phenomena, which concentrate  only on small  sections of the  observed
behavior. Experiments are often performed, guided by existing theories
to mainly  confirm certain  features predicted  by them,  rather  than
making independent measurements.  This process introduces some bias on
the  observations themselves, and  as a result,  features which can be
explained in a simpler way, may be  lost.  It is, therefore, necessary
to analyze and interpret the vast amount of experimental data which in
itself carries   a  lot of  information, e.g.   the  half-lives of all
decaying systems and  this can lead to   new interpretations based  on
greater generalities.

\section{ Duality in mass-time and Cantor theory:} 

In  this review,  a new  phenomenological  approach is  used  to unify
various types     of  experimental  measurements   to   understand the
fundamental particle and nuclear  aspects of physics.  Nuclear data of
decaying  systems  carry   information about  fundamental interactions
through their emitted  radiations like $\alpha$-,  $\beta$- particles.
Using only the masses  and lifetimes of decaying  systems, it is shown
that the observed data can be fitted to a unified scheme based on some
very fundamental considerations.  This  is done by  correlating masses
and lifetimes corresponding to different kinds of nuclear and particle
systems through an equation of the form [2] 

\begin{equation}
\frac{n}{2^n} = \frac{\hbar}{MT} = \frac{\Gamma}{M} 
\end{equation}

where h is the Planck's constant divided by $2\pi$. $M$ and $T$ in the
case of  fundamental particles  are the  masses and mean  lifetimes of
decaying particles. In  the cases of $\beta$- emitting radio-nuclides,
$M$  and  $T$ denote the neutron  mass  and  half-life of the decaying
nucleus.  M is taken to be the neutron  mass because the $\beta$-decay
process in a nucleus   can be effectively   reduced to the decay of  a
neutron or  a   proton   which  have almost  the   same   mass.    For
$\alpha$-emitting radio-nuclides,    $M$ and  $T$ are   chosen  as the
binding energy and half-life of the nucleus  because the whole nucleus
is involved in the process.  Using the experimental  values of $M$ and
$T$ [3,4], the quantity $n$ is  calculated. Various properties of this
crucial quantity are explained later.

The choice  of the particular equation (1)  comes about  from a deeper
basis  [5] provided  by  the Cantor's  Continuum theory which suggests
discreteness and continuity through  the concept of the denumerability
and non-denumerability by  comparison of the  cardinalities of various
sets, especially the   infinite ones [6].   One of   the principles of
Cantor theory says that  for a set  having `$n$' elements there always
corresponds   another   set    of $2^n$     elements  with  a   higher
cardinality. In  principle, the  difference in cardinalities  can give
rise to     `relative' discreteness  and   continuity.   The adjective
`relative' for  discreteness and continuity,  arises especially in the
context  of  physical quantities.    In  reality, we cannot  regard  a
physical quantity as absolutely  discrete.  The extent of  elements of
discreteness and continuity  depend on the  topology of the dimensions
in the space-time  domain of observation.   Any discrete quantity  can
always be transformed   into   a continuous  one  by   the  process of
interpolation [7].  Hence in  principle,  there would always  be  some
degree  of continuity  involved in  the discreteness.   The concept of
wave-particle duality which  says  that for each  (discrete)  particle
there is associated a (continuous) wave,  also supports this argument.
In  other  words,   we   shall    be   using  here  the   notion    of
discreteness-continuity in preference to particle-wave concept because
of  its more universal   behaviour.  It is   seen that  using  a wider
interpretation in  terms of discreteness-continuity motivated from the
theorems of  Cantor,  one is able  to  cover a lot  of  data under one
scheme.

Following the argument of the above paragraph, the width ($\Gamma$) or
the spread  of a line can be  associated with a more  continuity (thus
the higher cardinality) than  that of the  energy ($E$) of the levels.
In  other words, the lifetime   of a system  can  be associated with a
higher cardinality than that of its mass $ M$.  In this way, the ratio
$\frac{\Gamma}{M}$  ($\frac{\hbar}{MT}$) which suggests  a duality  in
mass and  lifetime  can be   related  to $\frac{n}{2^n}$ as  shown  in
eq. (1).

The  quantity $n$ in  eq.  (1) can be taken  as having integral values
[2]. This can be easily seen  to be the case  from Table 1 which gives
the deviation of the precise value  of $n$ from their nearest integral
values for elementary particles.  The  use of integral values of '$n$'
makes our analysis  more transparent to study  and interpret the data.
The  integralization  of `$n$' in  eq.(1)  suggests   some  sort of  a
discreteness in the $\frac{\Gamma}{M}$ ratios, a  result which is also
proved   within the more conventional  theory,  the `Quark Model' (for
details  see ref. [8]).  The differences in the   values of `$n$'  and
`$n_{exact}$' could arise from the  experimental errors of the widths.
However, a plot of $\frac{n-n_{exact}}{n}$  agains `$n$' clearly shows
that the distribution for $n= 6, 10, \cdots $ have a special structure
with many points with small differences.

The straight line  plot of $\log\left(\frac{\hbar}{MT}\right)$  versus
`$n$'  for elementary  particles  and $\alpha$- and $\beta$-  emitters
shown in Fig. 1, [2] clearly shows  the sectorwise distribution of the
integer `$n$' for different types of interactions with the presence of
forbidden regions in-between. Thus the  range of values of `$n$' turns
out to be  an   indicator of strength    of interaction in  the  decay
process. The  systems decaying through strong  interactions correspond
to values of $n < 10$. The systems which decay through electromagnetic
interactions  have values of `$n$' lying  between $10$ and $40$, while
those decaying through  weak forces have  values of  $n  > 40$.  As  a
result, most of the resonance particles lie in the region of $n < 10$,
while only  a few lie in the  region of higher `$n$'.   The $\alpha$-,
$\beta$- emitting radio-nuclides  have values of  $n > 40$  because of
their weak interacting nature.

We now consider the quantity 

\begin{equation}
p_o = - \log\left(\frac{n}{2^n}\right) 
\end{equation}

where `$n$' is an integer.  Table 2 gives the values of `$n$' when $p$
is very close to integers. 

\begin{center}
{\large {\bf Table 2}}
\vspace*{0.5cm}

\begin{footnotesize}
\begin{tabular}{rcccccccccccc}
n & = & 6 & 10 & 14 & 21 & 28 & 35 & 42 & 49 & 52 & 59 & 66\\[0.3cm]
$p_o$& = & 1.028&2.010&3.068&4.999&6.982&8.992&11.020&13.060&13.938&15.990&18.048\\[0.3cm]
p&=&1&2&3&5&7&9&11&13&14&16&18\\[0.3cm]
$p_o$-p&=&.028&.010&.068&-.001&-.018&-.008&.020&.060&-.062&-.010&.048\\[0.3cm]
\end{tabular}
\end{footnotesize}

\end{center}

It is seen that for  the elementary particle  region of $n < 65$,  the
quantity $p_o$ turns out to be close to prime integers for every `$n$'
being a multiple of $7$.  However the first two cases of $n = 6$, $10$
corresponding  to  the  resonance particles    are  exception to  this
rule. But this   rule does not  hold strictly  for the $\alpha$-   and
$\beta$-  emitters    with higher values  of    `$n$'.  By taking into
consideration only those values  of  `$n$' for which $p_o$-values  are
close to integers, amounts to a higher degeneracy which corresponds to
diluting the mass effects, thus resulting in quantization of lifetimes
as seen by Mac Gregor's work [9].

Now  we  shall study separately   different  sectors viz., fundamental
particles,  $\alpha$ and $\beta$ -    emitting radio-nuclides in   the
context of the present formalism motivated by Cantor's theory. 

\section{ Elementary Particles and the `$n$'-theory: }

The plot of `$n$'-values against  various elementary particles shows a
non-uniform  distribution.  It is seen   that in the   region of `$n$'
between  $24$ and $44$, no  particle except  the  $\pi^o$ is found. An
examination  of  the decay modes   of various  particles classified in
terms of `$n$'   shows that  the  majority  of resonances  that  decay
through  strong interactions have values  of $n < 10$.  Those decaying
through  electromagnetic   interactions have   values  of `$n$'  lying
between $10$ to $30$,  and those that  decay through weak interactions
have values of $n > 40$. 

If a toponium (ttbar)  is formed having mass around  $2 m_t  \sim 350$
GeV, and   it decays through hidden   strong interactions in  a manner
similar to bb   bar  and ccbar states,  then   its lifetime range   is
predicted to  be $2 \times  10^{-25} -  2.5 \times  10^{-21}$  s.  The
limits  of  $n$-values for the   top quark which  decays  through weak
interaction can be   set out as $40   < n$(top)$< 64$  considering the
maximum value of `$n$' to be $64$  for elementary particles. Thus, the
lifetime of the top quark with mass $174$ GeV [10] can be predicted to
lie between $10^{-16}$ s and $10^{-9}$ s. 

The  analysis of elementary  particles  in  terms of  the masses   and
lifetimes brings  out many new general  features in  the spectrum (see
Fig. 2). The  masses of the  unstable particles detected so  far range
from $0.1$ GeV to about $15$  GeV (excluding gauge bosons $W^{\pm}$, $
Z^o$) wherein  the most populated region corresponds  to $0.4$ - $2.5$
GeV. In  contrast to the short  mass  domain, the lifetime  domain for
unstable particles extends widely from  $10^{-24}$ s to $10^3$ s, with
several distinct peaks and regions of total absence of particles.  The
most prominent peak corresponds  to the resonance region of $10^{-24}$
s  to $10^{-20}$  s with  subsidiary peaks  at $10^{-13}$, $10^{-12}$,
$10^{-10}$, $10^{-9}$ s with a paucity of particles between $10^{-20}$
-  $10^{-13}$ s.  In  the entire region $2 \times  10^{-16} - 10^3$ s,
only one particle, the neutron is found. 

The identification of flavours of elementary particles can be achieved
by studying the plot of  log $T$ vs.  `$n$'(see  Fig.  3).  This  plot
shows some parallel  lines that correspond to  a constant value of $M$
which   in turn indicates   signature  of the presence of  subhadronic
particles.  Particles having similar substructure  lie on one straight
line associated with a constant value of mass corresponding to that of
constituent particles taken from conventional  Quark Model [11].  This
implies that  information  about  flavour substructure   of elementary
particles   is  hidden in  the  mass-lifetime   relation.   Some other
important properties  can be noted by  studying plot of  `$n$' against
different  flavours  of elementary particles. It   is seen from Fig. 4
that while  for light elementary   particles (unflavoured ones)  there
exists  large number of  resonances for  each  particle, but for heavy
particles like charmed strange baryons, bottom and charmed mesons very
few resonances have been discovered so far.  This calls for efforts to
detect some more new resonances for the heavy particles.  In contrast,
there are particles mostly quarkonia, e.g. cc bar and bb bar for which
only short  lived resonances are present  with no corresponding longer
lived particles in  the current experimental scenario.  This indicates
the unstable flavour antiflavour substructure of such particles.

\subsection{ `$n$'-Symmetries for elementary particles : }

Now we shall study some relations between the `$n$'-values ($\Gamma/M$
ratios) for the particles which are associated through (unitary) SU(3)
isospin symmetry.  Some striking  regularities  in the differences  in
their $n$-values are found [12]  for pairs of particles having similar
set of isospin quantum  numbers. The identification  of such pairs can
be easily facilitated  by the following  table which gives  the set of
mesonic and baryonic supermultiplets for SU(3) octets and singlets. 

\begin{center}

{\large{\bf Table 3}}
\vspace*{0.3cm}

\underline{Meson Supermultiplets}
\vspace*{0.3cm}

\begin{tabular}{cccc}
	   & Octets    &           &  Singlets   \\
triplet	   & singlet   &  doublet  &             \\
$\pi$	   &  $\eta$   &     K     &  $\eta$'    \\
$\rho$	   &  $\phi$   &     K     &  $\omega$   \\
  a        &    f'     &     K*    &     f
\end{tabular}
\vspace*{0.3cm}

\underline{Baryon Supermultiplets}
\vspace*{0.3cm}

\begin{tabular}{cccc}
           &  Octets   &           & Singlets \\
triplet	   &  singlet  &  doublet  &          \\
$\Sigma$   & $\Lambda$ &   N, $\Xi$    & $\Lambda$'
\end{tabular}

\end{center}

The `$n$'-values for  isospin   singlets and  neutral  member  of  the
isospin triplets belonging to SU(3) octets are found  to be related to
each other  through some periodicity. The  following equation (i) show
that   ($\pi^o$,$\eta$)     for  psuedoscalar      mesonic      octet,
($\rho^o$,$\omega$)    for      vector    mesonic         octet    and
($\Sigma^o$,$\Lambda$) for $1/2^+$ baryonic octet are examples of such
pairs. These  pairs  of states with  different  values of full isospin
(I), but  the same values   of its third  component $I_3$  ($=0$) show
following regularities:

(i)$ n_{\pi^o} - n_{\eta} = 2^2 + 1$ ; $ n_{\rho^o} - n_{\omega} = 2^2
   + 1$ ; $ n_{f1'} - n_{a1} = 2^2+1$ ; $ n_{\Lambda} - n_{\Sigma^o} =
   2^5 + 1$ 

		On  the  other hand,  the  singlets belonging to SU(3)
octets and their orthogonal singlets with same values of $I$ and $I_3$
($=0$),    show a   slightly    different  kind    of   periodicity in
$n$-values. Examples  of    such     pairs  for the      mesons    are
($\phi$,$\omega$), ($\eta$,$\eta$'), ($f_2$,$f_2$').  For the baryons,
the   excited states     $D_{03}    \Lambda$($1690$)  and      $D_{03}
\Lambda$($1520$) represent such a pair  because the ground state SU(3)
flavour singlet $\Lambda$ is forbidden by Fermi statistics. 

(ii)  $ n_{\phi} - n_{\omega}  =  2^1$ and  their higher excitations $
     n_{\phi 3} - n_{\omega 3} = 2^1$ 

\phantom{(ii)}  $ n_{\eta} - n_{\eta  '} = 2^3$  ; $n_{f2'} - n_{f2} =
 2^1$ ; $ n_{\Lambda (1520)} - n_{\Lambda (1690)} = 2^1 $ 

Two symmetries described above can be combined together to give a more
general result for association of neutral particles belonging to SU(3)
symmetry with same values of $I_3$ through the relation 

\begin{equation}
|\bigtriangleup n| = 2^q + | \bigtriangleup I | 
\end{equation}

where  $\bigtriangleup   n$, and    $\bigtriangleup I$  represent  the
differences in their  integer `$n$'  and isospin values  respectively.
It is to be noted that this relation holds for only integral values of
$\bigtriangleup I$.  Some   interesting properties of the integer  are
pointed out later. 

The  regularities found above in the  $n$-difference point  out to the
quantization of the  $\Gamma /M$ ratios  for the isospin levels  which
can be written in the mathematical form as [12] 

\begin{equation}
\left[\frac{MT}{\hbar}\right]_{n+\alpha} =  k
\left[\frac{MT}{\hbar}\right]_n 
\end{equation}

where subscripts in above equation denote the corresponding $n$-values
for $\frac{MT}{\hbar}$. Note that `$n$' and $\alpha$ are integers, and
$\alpha$   is the index used  for  the isospin.  This  equation can be
simplified by using eq. (1) to 

\begin{equation}
k = \left(\frac{n}{ n+\alpha}\right) 2^{\alpha} 
\end{equation}

The quantization of the isospin levels  suggests that $k$ should be an
integer  which is true   when $n + \alpha  =   2^q$, for every $q  \le
\alpha$. The second condition for $k$ to be an integer can be obtained
in the special case of `$n$' being a multiple of $\alpha$.  This gives
rise to several sets, as shown in Table 4. 
\begin{center}

{\large{\bf Table 4 }} 
\vspace*{0.3cm} 

\begin{tabular}{ccccccc}
        & & A & B & C & D & E \\ $\alpha$& = & n &  n/3 & n/7 & n/15 &
n/31   \\  k &    =  &$2^{n-1}$&  $3*2^{(n-6)/3}$   & $7*2^{(n-21)/7}$
&$15*2^{(n-60)/15}$ & $31*2^{(n-155)/31}$ \\ & & $n \ge  1$ & $n\ge 6$
& $ n \ge 21$ & $ n \ge 60$ & $ n \ge 155 $ 
\end{tabular}
\end{center}
\vspace*{0.3cm}

The  third row in the table  which gives the limits  of `$n$' for each
case, rules out the condition of $n = 0$ for the stable systems.  Thus
$ n  \rightarrow \infty $  is the only limit  for such systems.  It is
interesting to find that the region corresponding  to the sets B and C
matches fairly well with the  resonance region limits of  $6 \le n \le
24$, while the limits $ 2*21 \le n \le 60 $ derived  from sets C and D
agree well with the more stable region of $44 \le n  \le 64$. Thus the
assumption of quantization of isospin levels yields as seen from above
the results which   are in conformity with  the   observed ones.  This
equation applies only to particles  having isospin, but it does indeed
give  the required limits  for   various types  of  interactions  with
intermediate  `$n$'   for   non-isospin   particles  for  which   this
quantization  rule would  have  no  meaning.  These  are  some of  the
aspects which should be looked into.

As  stated earlier, the integer $\alpha$  is the isospin  index, or in
other words it is a measure of the spacing of  isospin levels. $I = 0$
is   a single level    while  $I  =  1/2$    is a  doubly   degenerate
level. Similarly  $I = 1$  has three-fold degeneracy  and  so on.  The
integer $\alpha$ has  different values for  different isospin doublets
like nucleon, kaon  etc., and also  for isospin triplets  like sigmas,
pions  etc. This suggests that   integer $\alpha$ apart from being  an
isospin index  can also be  distinguished by other parameters  such as
flavours unknown at present.

Based on the  general principles of the  tendency for a system to take
lowest energy shift intervals, the value of $\alpha$ must be lowest and
always positive. In  this  respect `$n$' represents the  lowest ground
state in terms of the isospin for each $I =  0$ particle. Now $\alpha$
will be the same  for  all members  of  an isospin multiplet.  On  the
other hand, the integer quantity $q$ would change for different values
of the  third  component ($I_3$)  of the  isospin which  distinguishes
various members of the  same isospin multiplet.   In particular  for a
triplet, q increases by one  unit as we move from  ($I_3 = 0$) neutral
member to the  ($|I_3| = 1$)  charged one .   The second case  (ii) of
same $I$, $I_3$ correspond to same  values of $\alpha$'s for each $n$,
so that their difference is equal to $2^m$,$ m$ being  an integer.  On
the other hand,  for case  (i)  which relates states having  different
$I$'s   with $\alpha$ being different,  a  different  kind of symmetry
exists.

\subsection{ Proton Life-time : }

We shall  now use the above discussion   for relating the `$n$'-values
for  the case  of different  members  of the  isospin multiplet  which
differ from each other only through  the third component $I_3$.  Since
$I$ is  same,  all members  have similar value   of $\alpha$.  But the
value  of  $q$   changes in  units  of   one.  This implies  that  the
$n$-difference is  equal to $2^q$. But  exciting part in this scenario
is that  the quantity $q$ can be  related to one  of the `$n$'-values,
and hence  the other  `$n$'-value   can be predicted.   This  we shall
explain by  taking up  various   isospin multiplets  available in  the
experimental data. 

\begin{itemize}
\item[1)]   We first    consider  the    isospin triplet  of    baryon
($\Sigma^{\pm}$, $\Sigma^{0}$). \\ For $\Sigma^{\pm}$ with $|I_3| = 1$
and $n = 54$, the postulate of choosing the  minimum value of $\alpha$
gives $q = 6$, and thus $\alpha  = 64 - 54=10$.  This implies that for
$\Sigma^0$ ($I_3=0$)  for which $\bigtriangleup I_3  = -1 q$ goes to $
q-1  = 5$ ,  while the  $\alpha$-value  remains the  same since we are
considering  the same multiplet.   Thus, we get $\alpha  =  2^5 - 10 =
22$. This  value of  `$n$' for  $\Sigma^0$  matches exactly with  that
obtained using experimental values.  Note that the difference in their
$n$-values is $n_{\Sigma\, \pm} - n_{\Sigma\, 0} = 2^5$.

\item[2)] Now, we take the example of  isospin triplet of psuedoscalar
mesons ($\pi^{\pm}$, $\pi^0$). For $\pi^{\pm}|I3 | =1$ and `$n$'$= 58$
which implies that minimum value of $\alpha$ is possible only for $q =
6$, which is  equal to $6$ ($64-58$).  With this value,  the $n$-value
for $\pi^0$ with $I_3=0$ and  $q = 6-1=5$,  is equal to $26$ ($32-6$).
This   value  of `$n$'  corresponds   to a life-time  of  $1.26 \times
10^{-17}$    s  for  $\pi^0$   in   comparison  to  its   experimental
$\mbox{value}^4$ of $8.4 \pm 0.6 \times 10^{-17}$ (with $n = 29$). 

\item[3)]  Encouraged  by  the  above  cases   which give fairly  good
agreement with the  experimental results, we  venture further into the
calculation  of  the  value of  `$n$'  for the    proton by using  the
corresponding value  for `$n$'  of  its   other doublet  member,   the
neutron.  Since $I_3=1/2$ for  proton and $I_3=-1/2$ for  the neutron,
so that $\bigtriangleup  I_3=1$ as we go from  neutron to proton.  For
the case of  neutron, $n=97$, thus  $\alpha =  2^7 - 97  = 31$.   As a
result, for the  corresponding case of the  proton for which $\alpha =
31$ ; $q  = 7+1$, $n$  is $2^{7+1} - 31  = 225$.   With this value  of
`$n$' for  the  proton, we get  the  lifetime of the  proton  as $5.33
\times  10^{33}$   years,  which   does  not   contradict  the present
experimental limit  of proton lifetime $>  10^{31} - 5 \times 10^{32}$
years [3, 12]. 

\end{itemize}

By analyzing relation (3) in context of three cases associating 

\begin{itemize}

\item[i)] neutral particles of triplet and singlet groups belonging to
an SU(3) octet 

\item[ii)] neutral particles of SU(3) octet and singlets 

\item[iii)] various members of isospin triplets (with different values
of I3), it is found  that the integer  q  takes values which are  only
prime numbers. e.g. 

$q   =1$ for  ($\phi$,   $\omega$), ($f_1$',  $a_1$);  $q   =  2$  for
 ($\pi^0$,$\eta$), ($\rho^o$,$\omega$), ($f_2$,$f_2$');  $q =  3$  for
 ($\eta$,$\eta$');  $q     =  5$    for  ($\pi^{\pm}$,$    \pi^0$)   ,
 ($\Sigma^{\pm}$, $\Sigma^0$)  ($\Sigma^0$,$\Lambda$) and $q  = 7$ for
 ($p$,$n$). 
\end{itemize}

It is intriguing to  find that the life-time of  a particle with known
mass can be  calculated using  that of  the  other particle  which  is
related  to the  former through  isospin  symmetry.  This  is  seen in
examples (1-3) which correspond to the particles with roughly the same
mass, and is   well supported by  cases  (i)  and (ii).   The  present
approach  does not  distinguish     between the particles  and   their
antiparticles with  respect  to their `$n$'-values.  This  formulation
holds true for all light-flavoured hadrons  which contain at least two
isospin flavours.   The  $\rho$-meson case   (with  $n  < 6$) is   not
considered  here  because in this case    of very short  life-time the
experimental errors are too large to  enable a unique determination of
integral $n$-value [2].   It is to  be noted that the present approach
does not work  for the systems consisting of  only one isospin flavour
(e.g.  kaons, bottom and charmed mesons) and of course for non-isospin
states.  Since there are not  many cases of mass multiplets  available
in the current experimental scenario[3], we don't yet have a more firm
basis for our formulation.   However, the limited number of multiplets
available support the hypothesis for a particular type of system.

The  regularities  described  above with  respect  to   `$n$'  are the
reflections of the symmetries in the life-times of the particles since
their  masses  play a passive  role  in the  $\Gamma/M$ ratios because
variations in the mass domain are very small  compared to those in the
life-times. This  suggests that the mean  life-time for the decay of a
particle  depends   on  its internal structure,     and  hence can  be
calculated inclusively.  This is also supported by  the  work of Mitra
and  Sharma [8].  It is   well known that  the decay  modes of various
particles are formulated by taking into consideration various internal
quantum numbers, e.g..   isospin,  strangeness, charm, etc.,   and are
subsequently  confirmed  by  experiments.  Therefore,  these  internal
quantum numbers  and their corresponding wave   functions must play an
important role in  deciding how a particle decays.   This calls for  a
detailed  analysis   of present  approach   in the  context   of  more
conventional theories. 

\section{ $\beta$-decay and `$n$' theory }

\subsection{ }

For the $\beta$-emitting nuclei since $M$ is  taken to be the constant
mass  of the neutron,  `$n$'  effectively represents  the behaviour of
halflife of the nucleus.  In the present case, the `$n$' values mostly
range from $70$ onwards   owing  to the   weak interacting  nature  of
$\beta$ decay (See Fig. 1).  But few light nuclei  have n values lying
in the region of strong interactions.   When $ Z/A$ is plotted against
integer $n$-values (see Fig.   5) the electron and   positron emitting
nuclei distinctly separate  out in the region  of `$n$' values smaller
than the  `$n$' for neutron.   This can be  explained by the fact that
electron emission results in  the increase  of  nuclear charge  by one
unit and  the positron emission   results in the decrease  of  nuclear
charge  by one  unit.   As a  result, the   upper branch  of the  plot
corresponds to  electron ($\beta^-$)   emitters  while the  lower  one
corresponds to the positron  ($\beta^+$)  emitters which is  just  the
reflection of the   former  curve.  This  plot   shows the exponential
behaviour of  $Z/A$ with respect  to `$n$', which  can be expressed as
$Z/A \sim (1+e^{-n})  / 2$  for $\beta^-$  emitters   and $ Z/A   \sim
(1-e^{-n})/2$ for $\beta^+$ emitters. 

The proportionality of the halflife of a  nucleus on the $Z/A$ seen in
the above paragraph,   suggests  the  dependence of  nuclear   binding
energies on the halflife through  $Z/A$.   The least square fit  curve
(Fig.    6a,  6b) of the binding    energies  shows that   they can be
expressed as [13]

\begin{equation}
\frac{BE}{A} = \alpha(M_n - M_p)
 \left\{\log{\left(\frac{Z_n}{n_0}\right)} - 0.316
 \left(\log{\frac{Z_n}{n_0}}\right)^2 \right\} 
\end{equation}

This formula holds  for all $\beta^-$,  $\beta^+$ emitters  as well as
  K-electron capture nuclei.  Since $\beta$-decay  of a nucleus can be
  effectively reduced to the decay of  a neutron inside a nucleus into
  a proton and an electron, the difference ($M_n  - M_p m_e) \sim (M_n
  -  M_p)$, between  their masses should   play an  important  role in
  contributing    to   the   binding energies.     $BE/A$    shows its
  proportionality to  the half-lives  through the logarithmic argument
  ($Z_n/n_0$)  where $n_0$ represents  the minimum values of `$n$' for
  each  case  which   is  equal   to $70$  and    $75$ for  $\beta^-$,
  $\beta^+$-emitters  respectively.   Here   we have  not taken   into
  account the nuclei  which  do not  decay through  weak  interaction.
  Since the  `$n$'  values  for  $\beta$-emitters range  from $70$  to
  $150$, the ratio  $n/n_0$  changes from  $1$ to  $2$.  As a  result,
  $Z_n/n_0$  moves from $Z$   to $A$ as  we go  towards nuclei  having
  longer half-lives.  The quantity $\alpha$  in the expression  $BE/A$
  represents a dimensionless normalising factor  of value $8.45$ which
  is same for  all cases.  Fig. (6a,  6b)  give the  comparison of the
  observed $BE/A$  and those calculated using  the  formula (14).  The
  agreement holds fairly   well  considering the huge   number ($1300$
  cases) of nuclei involved. 

Though only unstable systems are under consideration, it is clear from
Fig.  (7a \& 7b) that as $N$ and $Z$ approach magic numbers, the value
of  `$n$'   increases  indicating longer    lifetime  and  thus  great
stability.

The explanation of the $\beta$-spectrum  is given by Fermi's theory of
$\beta$-decay which expresses the halflife of decaying nucleus as [13] 

\begin{equation}
\frac{0.693}{T_{1/2}} = |M|^2 f\frac{\left(g^2 m e^2 c^4\right)}{2\pi^3 h^7}
\end{equation}

Here $|M|^2$ is  the  nuclear matrix element   which accounts for  the
effects  of a particular initial and  final nuclear states.  The Fermi
function f takes into account  the effect of the  coulomb field of the
nucleus.  $g$ is the weak interaction constant.   The product $f T$ is
called  the comparative halflife  which gives us a  way to compare the
$\beta$-decay probabilities  in different nuclei.   Since the quantity
$f$  incorporates the dependence on  the atomic number and the maximum
electron  energy, the difference in  $f T$ values  [15] must be due to
differences  in the nuclear matrix elements.   This is shown in Fig. 8
which gives the plot of $|M_{e1}|^2$ vs $n$.  Fig.  9 shows the effect
of incorporation of  function $f$  which results  in  the reduction of
$n$-values.  The histogram also  depicts the frequency distribution of
superallowed,   allowed and forbidden   transmission which  shows that
maximum number of nuclei correspond to the $\log{fT}$ values of 5.3 as
seen in Ref. [16]. In this  plot the reduced  n is a direct measure of
$\log{fT}$ values. 

\subsection{ Ground state spins of $\beta$-emitters and their half-lives : }

The groundstate spins of   $\beta$-emitters can be obtained  from  the
 same equation (2) i.e., $p_0 =  n\log{2} - \log{n}$.  Let us consider
 a situation   where $p_0$ is  not equal  to $n\log{2}-\log{n}$ but is
 modified by spin and many body effects to 

\begin{equation}
p' = n\log{ 2(q-1)} - \log{n} + \log{(Z/A)} 
\end{equation}

where $2$ is replaced by $2 (q-1)$ and has a term containing $Z/A$, as
$Z/A$ plays an important part in spin and other many body effects.  We
assume that ($q-1$)  is directly proportional to  spin and ($q-1$)  is
always  negative and  ($1-q$) is always   positive.   As we  shall see
later, these positive and  negative values result in space  distortion
of `$n$'.   Fig 10 shows the  effect of change  in $p'-p_0$ as ($q-1$)
departs from $1$ on either side for nuclides.   The set of nuclides on
the left corresponds  to those of negative parity  and vice versa.  It
can be noted that  the the two parity  distributions are not identical
and those with values of ($q-1$) being negative  have real values only
when `$n$' is even. 

The  departure from $\log{2}$ to   $\log{2(q-1)}$ implies expansion of
$n$--space geometry  for the -ve  components, while for +ve values the
relative values of  `$n$' are contracted.  One  can conjecture that it
is this change in geometry of the $n$-space that gives rise to spin. 

\section{ $\alpha$-decay and the $n$-theory:} 

The general features  of   $\alpha$- emission which occurs  mostly  in
heavy nuclei with mass number greater than $150$, can be accounted for
by  a    quantum mechanical   theory    developed  in   $1928$  almost
simultaneously by Gamow and Gurney and Condon.  The central feature of
this one-body model is that  the $\alpha$ particle is preformed inside
the   nucleus.  In fact  there  is  not  much reason  to believe  that
$\alpha$-particles do exist   separately  within the   heavy  nucleus;
nevertheless  the  theory works  quite  well, especially for even-even
nuclei.   The  success  of     the   theory  does    not   prove  that
$\alpha$-particles  are  preformed but merely they   behave as if they
were. The theory gives the possibility of  leakage or tunneling of the
$\alpha$-particle  through the large potential  barrier of the nucleus
which  is absolutely impossible in  the classical interpretation.  The
leakage  probability is  so small that  the  $\alpha$-particle  on the
average must make $10^{38}$  tries before it  escapes (this amounts to
$10^{21}$    per     sec !).   The     disintegration   constant of an
$\alpha$-emitter  is   given  in one body   theory   by the  following
expression [13]: 

\begin{equation}
\lambda = f P 
\end{equation}

where $f$ is the frequency with which  the particle presents itself at
the barrier, which is given by its velocity  in the nucleus divided by
the radius of  the nucleus.  $P$  is  the probability of  transmission
through the barrier which is given by the following expression 

\begin{equation}
P = e^{-2G} 
\end{equation}

The Gamow factor $G$ is [14] 

\begin{equation}
G = \left[ \frac{(2 m c^2)}{(\hbar c)^2}Q\right]^{1/2} \frac{(ZZ' e^2)}{(4\pi
 \epsilon_0)} \left[\frac{\pi}{2} \,-\,
 2\left(\frac{Q}{B}\right)^{1/2}
\right] 
\end{equation}

where $m$, $Z$ is the mass and charge of the $\alpha$-particle and $Q$
 is its  kinetic  energy. $B$  is the  binding energy  of the daughter
 nucleus and  $Z'$  is its  charge. So, the   final expression  can be
 written in terms of half-life as 

\begin{equation}
f = \left(\frac{0.693}{T_{1/2}}\right) e{^-2G} 
\end{equation}

If we  replace $T_{1/2}$  by the  binding  energies and the  $n$ using
equation (1) for the $\alpha$ particle 

\begin{displaymath}
\frac{n}{2^n} = \frac{\hbar}{M T_{1/2}} = \frac{\hbar}{V_0 T_{1/2}} 
\end{displaymath}

where $V_0$ is the binding energy of the parent nucleus, we get 

\begin{equation}
f = \left(\frac{0.693 V_0 n}{ \hbar 2^n} \right) e^{-2G} 
\end{equation}

Fig. 11 shows that  $f$ is constant over  a wide range  of $n$-values,
and is of the order of $10^{21}$ to $10^{22}$ per s [14].

\section{ Excited Nuclear Levels and Elementary Particles: }

The transition of a  ground state nucleus to  an excited one of higher
energy occurs under the  external influences when the  required energy
is transferred to  the nucleus through  interaction with  an energetic
particle. When  excited externally one  or many nucleons (depending on
the quantity of energy applied) occupy higher energy levels. Since the
nucleon  levels are separated  by finite energy intervals, the nucleus
cannot receive any quantity  of   energy but only  strictly   definite
portions   precisely corresponding   to    the energies   of   nucleon
transitions from lower to higher states.  In  this section, we propose
that the energy  is absorbed by the  nucleus in a  discrete way in the
form of virtual mesons mainly the light ones.  As a result, the system
gets excited to a level whose width and energy depend  on the mass and
decay time of the meson involved in  the process.  The type of virtual
meson depends on the external energy imparted to the nucleus. 

Each excited  level  of a  nucleus   is associated   to an  elementary
particle   which   is     identified    by  comparing    the     ratio
$\log{(E_r/\Gamma)}$ [17]   (Fig. 12) for  an  excited level of energy
$E_r$ and width $\Gamma$ with the $\log{(MT/\hbar)}$ for an elementary
particle of mass $M$ and lifetime  $T$.  In the  Fig. 12, integer $Dn$
is given  by    $p=\log{(2^{D_n}/Dn)}$.   The quantity  `$n$'   for  a
resonance level is calculated  from $E_r/\Gamma$ by using eq.(1) which
helps in the identification of the particle involved in the excitation
process (see Table 5). The  elementary particles involved are found to
be light mesons   [18].   Fig.  13 gives  the  number  distribution of
various $n$-values, which shows that it is  the $\omega$-meson that is
mostly involved in the excitation process.

\section{ Comparison of present formalism with other  appro\-aches: }

The present formalism [19] which deals with duality in particle masses
and  life-times  gives  a   discreteness in  $\Gamma/M$  ratios.  This
approach, of course, has not been attempted  for the first time. Nambu
as  early as 1952 had  written a paper   on empirical mass spectrum of
elementary particles  [20] wherein he  had suggested a discreteness of
particle masses by  taking electron mass as  the basic unit.  In other
work by Mac Gregor [9,21], the quantization  of particle lifetimes was
found in  terms of the muon  lifetime.  The present approach is unique
in the sense that it incorporates two  discrete quantities together in
a  unified framework without  choosing a  basis dependent on  electron
masses  or muon lifetimes.    Moreover the  present  approach is  more
transparent which can   bring out  many  more properties  in a  simple
manner.

Mac Gregor's model [9] which predicts quantization in the lifetimes of
the  elementary particles represents a higher  degree of degeneracy in
the quantized levels compared to the present approach.  As seen in the
present  text the  inclusion  of  particles  masses along  with  their
lifetimes thus suggesting a duality between  them brings out hyperfine
quantization for elementary particles. On the  other hand Mac Gregor's
approach  does not  incorporate the  particle   masses in his  formula
[9]. It can  be easily  seen that  the quantum levels  in his  formula
correspond to a  special case of present  formalism.  If  we constrain
the     quantity   $p_0$ to  take   only   integral   values, then the
corresponding $n$-values  obtained   (see Table  2)   have one to  one
correspondence with the quanta in Mac Gregor's approach [9].

\section{ Conclusion : }

It  is seen  that from the    parameter `$n$' or   `$p$'  based on the
life-time and  the energy available for  the decaying system,  one can
arrive at  several properties of    fundamental particles,  spin   and
parities of the $\beta$-emitters and the  number of times the $\alpha$
particle strikes the barrier of the radioactive  nuclides before it is
expelled.  The  Binding Energy of  about $1300$ nuclei can be obtained
by an empirical  formula  based on the  above  parameter.  It is  also
possible to   get the systematics  of the   resonance levels  of light
nuclei, as a  process in which  short-lived fundamental particles play
the role of a virtual particle for the transfer of energy. 

The relation  between this method based on   Cardinality and the usual
Quantum Mechanical  approach has yet to be  worked out.   The authors,
however, feel that there will  not be any  inconsistency as waves  and
particles are a special form of continuity and discreteness. 

We are grateful to Prof. B.V.   Sreekantan for many useful comments on
the paper.   We  also thank Dr.    (Mrs) Sundaramma Srikanta   and Mr.
K.S. Rama Krishna for their help in preparing the paper.

\newpage

\begin{center}
{\LARGE {\bf References }} 
\end{center}

\begin{itemize}

\item[1.] H.  A. Bethe and R.F.  Bacher, Nuclear Physics A: Stationary
States of Nuclei, Revs. of Mod. Phys. 8, 82 (1936). 

	H. A. Bethe, Nuclear Physics B: Nuclear Dynamics, Theoretical,
	Revs. of Mod. Phys. 9, 69 (1937). 

	M.   Stanley Livingston and  H.  A. Bethe, Nuclear Physics  C:
	Nuclear Dynamics,  Experimental,  Revs. of  Mod. Phys.  9, 245
	(1937). 

\item[2.] Raja Ramanna   and B.V. Sreekantan,  On the  Correlations of
Masses  and   Lifetimes of   Radionuclides and  Fundamental Particles,
Modern Physics Letter A10, 741 (1995). 

\item[3.]   Review  of  Particle  Properties, Phys.Rev.  D50,  Part I,
August (1994). 

\item[4.]  A.  H Wapstra and G.  Audi, The 1983 Atomic Mass Evaluation
(I): Atomic Mass Table, Nucl. Phys. A432, 1 (1985). 

\item[5.]  Raja Ramanna,  Concept of discreteness,  continuity and the
Cantor continuum  theory  as related to the   life-time and  masses of
elementary particles, Current Science, Vol 65, 472 (1993). 

\item[6.]    R.  Courant and  H. Robbins,   ``What is  Mathematics: An
Elementary Approach to Ideas and Methods'', Oxford Univ. Press, (1978)
pp 84. 

\item[7.] Patterns of Chaos 

\item[8.] A.N.Mitra and Anju Sharma, Intl. J. Mod. Phys. A11 (1996) 

\item[9.] M. H. Mac Gregor, IL Nuovo Cimento A103, 983 (1990). 

\item[10.] F. Abe et al Phys. Rev. D50, 2966 (1994). 

\item[11.] A. N. Mitra and R. Ramanathan, Role of Hadronic Dynamics in
Proton Decay, Z. Fur Phys. 22, 351 (1984) 

\item[12.] Anju   Sharma  and Raja  Ramanna,   Particle  Symmetries in
$\Gamma$/M Ratios and  the Lifetime of  Proton, Modern Physics Letters
A11, 2335 (1996). 

\item[13.] Kenneth  S.  Krane, ``Introductory Nuclear  Physics'', John
Wiley \& Sons. Inc.  (1988). 

\item[14.] Raja  Ramanna, Duality of Masses  and  Lifetimes in Quantum
Systems, Intl. J. Mod. Phys. A11, 5081(1996). 

\item[15.]  Arnold    M.  Feingold,  ``Table      of  ft  values    in
$\beta$-decay'', Rev. of Mod. Phys. 23, 10 (1951). 

\item[16.]  W. Meyer hof, Elements  of Nuclear  Phys. -New York McGraw
Hill (1967) 

\item[17.] Lauritsen T and  Ajzenberg-Selove F, Energy Levels of Light
Nuclei (VII). A=5-10, Nuclear Physics 78 (1966) 

\item[18.] Raja Ramanna, The Excited  States of Light Nuclei and their
Relation to Fundamental Particles, Submitted for publication. 

\item[19.] Raja Ramanna, Mass-Time  Duality and the Ground State Spins
of $\beta$-emitting nuclei, Submitted for publication. 

\item[20.] Y. Nambu, Prog. Theor. Phys. 29, 1091 (1990) 

\item[21.] D. Akers, IL Nuovo Cimento, A105 , 935 (1992). 
\end{itemize}

\newpage

\begin{center}
{\large{\bf       TABLE 1}}
\end{center}
\vspace*{0.3cm}

\begin{tabular}{rlrrrrr}
Sl.  &   Particle   &    Mass    &  Width    &$n_{exact}$ & $n$ &  $D_{n1}$\\
No.  &Identification&    (MeV)   &  (MeV)    &         &     & ($n-n_{exact}$)\\
  1  & $\pi^{\pm}$  &   1.40E+03 &  2.53E-13 &  58.15  &  58 &   0.15\\
  2  & $\pi^o$      &   1.35E+03 &  7.85E-05 &  28.88  &  29 &   0.12\\
  3  & $\eta$       &   5.47E+03 &  1.20E-02 &  23.34  &  23 &   0.34\\
  4  & $\rho$(770)  &   7.70E+03 &  1.51E+03 &   4.52  &   5 &   0.48\\
  5  & $\omega$(782)&   7.82E+03 &  8.43E+01 &   9.82  &  10 &   0.18\\
  6  &$\eta$(958)   &   9.58E+03 &  2.01E+00 &  16.23  &  16 &   0.23\\
  7  &$f_o$(980)    &   9.80E+03 &  2.20E+03 &   4.23  &   4 &   0.23\\
  8  & $a_o$(980)   &   9.82E+03 &  1.76E+03 &   4.71  &   5 &   0.29\\
  9  & $\phi$(1020) &   1.02E+04 &  4.43E+01 &  11.34  &  11 &   0.34\\
 10  & $h_1$(1170)  &   1.17E+04 &  3.60E+03 &   3.50  &   4 &   0.50\\
 11  & $b_1$(1235)  &   1.23E+04 &  1.42E+03 &   5.59  &   6 &   0.41\\
 12  & $a_1$(1260)  &   1.23E+04 &  4.00E+03 &   3.36  &   3 &   0.36\\
 13  & $f_2$(1270)  &   1.28E+04 &  1.85E+03 &   5.14  &   5 &   0.14\\
 14  & $f_1$(1285)  &   1.28E+04 &  2.40E+02 &   8.88  &   9 &   0.12\\
 15  &$\eta$(1295)  &   1.30E+04 &  5.30E+02 &   7.51  &   8 &   0.49\\
 16  &$f_0$(1300)   &   1.25E+04 &  2.75E+03 &   4.27  &   4 &   0.27\\
 17  &$\pi$(1300)   &   1.30E+04 &  4.00E+03 &   3.50  &   4 &   0.50\\
 18  &$a_2$(1320)   &   1.32E+04 &  1.07E+03 &   6.26  &   6 &   0.26\\
 19  &$f_1$(1420)   &   1.43E+04 &  5.20E+02 &   7.72  &   8 &   0.28\\
 20  &$\omega$(1420)&   1.42E+04 &  1.74E+03 &   5.47  &   5 &   0.47\\
 21  &$\eta$(1440)  &   1.42E+04 &  6.00E+02 &   7.46  &   7 &   0.46\\
 22  &$\rho$(1450)  &   1.47E+04 &  3.10E+03 &   4.36  &   4 &   0.36\\
 23  &$f_1$(1510)   &   1.51E+04 &  3.50E+02 &   8.52  &   9 &   0.48\\
 24  &$f_2$(1525)   &   1.53E+04 &  7.60E+02 &   7.16  &   7 &   0.16\\
 25  &$f_0$(1590)   &   1.58E+04 &  1.80E+03 &   5.62  &   6 &   0.38\\
 26  &$\omega$(1600)&   1.66E+04 &  2.80E+03 &   4.84  &   5 &   0.16\\
 27  &$\omega_3$(1670)  &   1.67E+04 &  1.73E+03 &   5.80  &   6 &   0.20\\
 28  &$\pi_2$(1670) &   1.67E+04 &  2.40E+03 &   5.16  &   5 &   0.16\\
 29  &$\phi$(1680)  &   1.68E+04 &  1.50E+03 &   6.08  &   6 &   0.08\\
 30  &$\rho_3$(1690)&   1.69E+04 &  2.15E+03 &   5.40  &   5 &   0.40\\
 31  &$\rho$(1700)  &   1.70E+04 &  2.35E+03 &   5.24  &   5 &   0.24\\
 32  &$f_1$(1710)   &   1.71E+04 &  1.40E+03 &   6.25  &   6 &   0.25\\
 33  &$\phi_3$(1850)&   1.85E+04 &  8.70E+02 &   7.27  &   7 &   0.27\\
 34  &$f_2$(2010)   &   2.01E+04 &  2.02E+03 &   5.86  &   6 &   0.14\\
 35  &$f_4$(2050)   &   2.04E+04 &  2.08E+03 &   5.83  &   6 &   0.17\\
\end{tabular}

\begin{tabular}{rlrrrrr}
Sl.  &   Particle   &    Mass    &  Width    &$n_{exact}$ & $n$ &  $D_{n1}$\\
No.  &Identification&    (MeV)   &  (MeV)    &         &     & ($n-n_{exact}$)\\
 36  &$f_2$(2300)   &   2.30E+04 &  1.49E+03 &   6.68  &   7 &   0.32\\
 37  &$f_2$(2340)   &   2.34E+04 &  3.19E+03 &   5.26  &   5 &   0.26\\
 38  &$K^{\pm}$     &   4.94E+03 &  5.33E-13 &  58.91  &  59 &   0.09\\
 39  &$K^0_S$       &   4.98E+03 &  7.38E-11 &  51.62  &  52 &   0.38\\
 40  &$K^0_L$       &   4.98E+03 &  1.27E-13 &  61.04  &  61 &   0.04\\
 41  &$K^*$(892)$^{\pm}$ &   8.92E+03 &  4.98E+02 &   6.95  &   7 &   0.05\\
 42  &$K^*$(892)$^0$    &   8.96E+03 &  5.05E+02 &   6.94  &   7 &   0.06\\
 43  &$K_1$(1270)   &   1.27E+04 &  9.00E+02 &   6.52  &   7 &   0.48\\
 44  &$K_1$(1400)   &   1.40E+04 &  1.74E+03 &   5.45  &   5 &   0.45\\
 45  &$K^*$(1410)   &   1.41E+04 &  2.27E+03 &   4.93  &   5 &   0.07\\
 46  &$K_0^*$(1430) &   1.43E+04 &  2.87E+03 &   4.47  &   4 &   0.47\\
 47  &$K_2^*$(1430)$^{\pm}$ &   1.43E+04 &  9.84E+02 &   6.56  &   7 &   0.44\\
 48  &$K_2^*$(1430)$^o$ &   1.43E+04 &  1.09E+03 &   6.38  &   6 &   0.38\\
 49  &$K^*$(1680)   &   1.71E+04 &  3.23E+03 &   4.60  &   5 &   0.40\\
 50  &$K_2$(1770)   &   1.77E+04 &  1.86E+03 &   5.78  &   6 &   0.22\\
 51  &$K_3^*$(1780) &   1.77E+04 &  1.64E+03 &   6.01  &   6 &   0.01\\
 52  &$K_2$(1820)   &   1.82E+04 &  2.76E+03 &   5.05  &   5 &   0.05\\
 53  &$K_4^*$(2045) &   2.05E+04 &  1.98E+03 &   5.93  &   6 &   0.07\\
 54  &$D^{\pm}$     &   1.87E+04 &  6.24E-09 &  46.99  &  47 &   0.01\\
 55  &$D^0$         &   1.86E+04 &  1.59E-08 &  45.60  &  46 &   0.40\\
 56  &$D^*$(2007)$^0$ &   2.01E+04 &  2.10E+01 &  13.67  &  14 &   0.33\\
 57  &$D^*$(2010)$^{\pm}$ &   2.01E+04 &  1.31E+00 &  18.07  &  18 &   0.07\\
 58  &$D_1$(2420)$^0$ &   2.42E+04 &  1.80E+02 &  10.45  &  10 &   0.45\\
 59  &$D_2^*$(2460)$^0$ &   2.46E+04 &  2.10E+02 &  10.22  &  10 &   0.22\\
 60  &$D_2^*$(2460)$^\pm$     &   2.46E+04 &  2.30E+02 &  10.06  &  10 &   0.06\\
 61  &$D_s^{\pm}$ &   1.97E+04 &  1.41E-08 &  45.86  &  46 &   0.14\\
 62  &$D_s^{* \pm}$      &   2.11E+04 &  4.50E+01 &  12.51  &  13 &   0.49\\
 63  &$D_{s1}$(2536)$^{\pm}$      &   2.54E+04 &  2.30E+01 &  13.90  &  14 &   0.10\\
 64  &$B^{\pm}$      &   5.28E+04 &  4.28E-09 &  49.10  &  49 &   0.10\\
 65  &$B^0$      &   5.28E+04 &  4.39E-09 &  49.06  &  49 &   0.06\\
 66  &$B^0_S$      &   5.38E+04 &  4.92E-09 &  48.92  &  49 &   0.08\\
 67  &$\eta_c(1S)$      &   2.98E+04 &  1.03E+02 &  11.72  &  12 &   0.28\\
 68  &$J/\psi$(1S)      &   3.10E+04 &  8.80E-01 &  19.37  &  19 &   0.37\\
 69  &$\chi^{c0}$1P      &   3.42E+04 &  1.40E+02 &  11.44  &  11 &   0.44\\
 70  &$\chi^{c1}$1P      &   3.51E+04 &  8.80E+00 &  15.95  &  16 &   0.05\\
\end{tabular}

\begin{tabular}{rlrrrrr}
Sl.  &   Particle   &    Mass    &  Width    &$n_{exact}$ & $n$ &  $D_{n1}$\\
No.  &Identification&    (MeV)   &  (MeV)    &         &     & ($n-n_{exact}$)\\
 71  &$\chi^{c2}$1P      &   3.56E+04 &  2.00E+01 &  14.66  &  15 &   0.34\\
 72  &$\psi$(2S)      &   3.69E+04 &  2.77E+00 &  17.85  &  18 &   0.15\\
 73  &$\psi$(3770)      &   3.77E+04 &  2.36E+02 &  10.74  &  11 &   0.26\\
 74  &$\psi$(4040)      &   4.04E+04 &  5.20E+02 &   9.52  &  10 &   0.48\\
 75  &$\psi$(4160)      &   4.16E+04 &  7.80E+02 &   8.88  &   9 &   0.12\\
 76  &$\psi$(4415)      &   4.42E+04 &  4.30E+02 &  10.00  &  10 &   0.00\\
 77  &$\gamma$(1S)      &   9.46E+04 &  5.25E-01 &  21.91  &  22 &   0.09\\
 78  &$\gamma$(2S)      &   1.00E+05 &  4.40E-01 &  22.27  &  22 &   0.27\\
 79  &$\gamma$(3S)      &   1.04E+05 &  2.63E-01 &  23.11  &  23 &   0.11\\
 80  &$\gamma$(4S)      &   1.06E+05 &  2.38E+02 &  12.42  &  12 &   0.42\\
 81  &$\gamma$(10860)      &   1.09E+05 &  1.10E+03 &   9.93  &  10 &   0.07\\
 82  &$ \gamma$(11020)     &   1.10E+05 &  7.90E+02 &  10.51  &  11 &   0.49\\
 83  &$n$      &   9.40E+03 &  7.43E-24 &  96.62  &  97 &   0.38\\
 84  &$N$(1440)$P_{11}$      &   1.45E+04 &  3.50E+03 &   4.07  &   4 &   0.07\\
 85  &$N$(1520)$D_{13}$      &   1.52E+04 &  1.23E+03 &   6.28  &   6 &   0.28\\
 86  &$N$(1535)$S_{11}$      &   1.54E+04 &  1.75E+03 &   5.62  &   6 &   0.38\\
 87  &$N$(1650)$S_{11}$      &   1.66E+04 &  1.68E+03 &   5.85  &   6 &   0.15\\
 88  &$N$(1675)$D_{15}$      &   1.68E+04 &  1.60E+03 &   5.96  &   6 &   0.04\\
 89  &$N$(1680)$F_{15}$      &   1.68E+04 &  1.30E+03 &   6.35  &   6 &   0.35\\
 90  &$N$(1700)$D_{13}$      &   1.70E+04 &  1.00E+03 &   6.86  &   7 &   0.14\\
 91  &$N$(1710)$P_{11}$      &   1.71E+04 &  1.50E+03 &   6.12  &   6 &   0.12\\
 92  &$N$(1720)$P_{13}$      &   1.72E+04 &  1.50E+03 &   6.13  &   6 &   0.13\\
 93  &$N$(2190)$G_{17}$      &   2.15E+04 &  4.50E+03 &   4.38  &   4 &   0.38\\
 94  &$N$(2220)$H_{19}$      &   2.25E+04 &  4.35E+03 &   4.54  &   5 &   0.46\\
 95  &$N$(2250)$G_{19}$      &   2.24E+04 &  3.80E+03 &   4.82  &   5 &   0.18\\
 96  &$N$(2600)$I_{1,11}$    &   2.65E+04 &  6.50E+03 &   4.03  &   4 &   0.03\\
 97  &$\bigtriangleup$(1232)$P_{33}$      &   1.23E+04 &  1.20E+03 &   5.92  &   6 &   0.08\\
 98  &$\bigtriangleup$(1600)$P_{33}$      &   1.63E+04 &  3.50E+03 &   4.32  &   4 &   0.32\\
 99  &$\bigtriangleup$(1620)$S_{31}$      &   1.65E+04 &  1.50E+03 &   6.04  &   6 &   0.04\\
100  &$\bigtriangleup$(1700)$D_{33}$      &   1.72E+04 &  3.00E+03 &   4.76  &   5 &   0.24\\
101  &$\bigtriangleup$(1900)$S_{31}$      &   1.90E+04 &  1.90E+03 &   5.87  &   6 &   0.13\\
102  &$\bigtriangleup$(1905)$F_{35}$      &   1.90E+04 &  3.60E+03 &   4.59  &   5 &   0.41\\
103  &$\bigtriangleup$(1910)$P_{31}$      &   1.90E+04 &  2.30E+03 &   5.49  &   5 &   0.49\\
104  &$\bigtriangleup$(1920)$P_{33}$      &   1.94E+04 &  2.25E+03 &   5.58  &   6 &   0.42\\
105  &$\bigtriangleup$(1930)$D_{35}$      &   1.95E+04 &  3.50E+03 &   4.70  &   5 &   0.30\\
\end{tabular}

\begin{tabular}{rlrrrrr}
Sl.  &   Particle   &    Mass    &  Width    &$n_{exact}$ & $n$ &  $D_{n1}$\\
No.  &Identification&    (MeV)   &  (MeV)    &         &     & ($n-n_{exact}$)\\
106  &$\bigtriangleup$(1950)$F_{3^-}$     &   1.95E+04 &  3.20E+03 &   4.89  &   5 &   0.11\\
107  &$\bigtriangleup$(2420)$H_{3.11}$    &   2.40E+04 &  4.00E+03 &   4.86  &   5 &   0.14\\
108  &$\Lambda$      &   1.12E+04 &  2.50E-11 &  54.42  &  54 &   0.42\\
109  &$\Lambda$(1405)$S_{01}$      &   1.41E+04 &  5.00E+02 &   7.76  &   8 &   0.24\\
110  &$\Lambda$(1520)$D_{03}$      &   1.52E+04 &  1.56E+02 &   9.91  &  10 &   0.09\\
111  &$\Lambda$(1600)$P_{01}$      &   1.63E+04 &  1.50E+03 &   6.03  &   6 &   0.03\\
112  &$\Lambda$(1670)$S_{01}$      &   1.67E+04 &  3.75E+02 &   8.57  &   9 &   0.43\\
113  &$\Lambda$(1690)$D_{03}$      &   1.69E+04 &  6.00E+02 &   7.77  &   8 &   0.23\\
114  &$\Lambda$(1800)$S_{01}$      &   1.79E+04 &  3.00E+03 &   4.84  &   5 &   0.16\\
115  &$\Lambda$(1810)$P_{01}$      &   1.80E+04 &  1.50E+03 &   6.21  &   6 &   0.21\\
116  &$\Lambda$(1820)$F_{05}$      &   1.82E+04 &  8.00E+02 &   7.39  &   7 &   0.39\\
117  &$\Lambda$(1830)$D_{05}$      &   1.82E+04 &  8.50E+02 &   7.28  &   7 &   0.28\\
118  &$\Lambda$(1890)$P_{03}$      &   1.88E+04 &  1.30E+03 &   6.56  &   7 &   0.44\\
119  &$\Lambda$(2100)$G_{07}$      &   2.10E+04 &  1.75E+03 &   6.21  &   6 &   0.21\\
120  &$\Lambda$(2110)$F_{05}$      &   2.12E+04 &  2.00E+03 &   5.97  &   6 &   0.03\\
121  &$\Lambda$(2350)$H_{09}$      &   2.36E+04 &  1.75E+03 &   6.43  &   6 &   0.43\\
122  &$\Sigma^+$    &   1.19E+04 &  8.25E-11 &  52.75  &  53 &   0.25\\
123  &$\Sigma^0$    &   1.19E+04 &  8.91E-02 &  21.45  &  21 &   0.45\\
124  &$\Sigma^-$    &   1.20E+04 &  4.46E-11 &  53.67  &  54 &   0.33\\
125  &$\Sigma$(1385)$P_{13}^+$    &   1.38E+04 &  3.58E+02 &   8.32  &   8 &   0.32\\
126  &$\Sigma$(1385)$P_{13}^0$    &   1.38E+04 &  3.60E+02 &   8.31  &   8 &   0.31\\
127  &$\Sigma$(1385)$P_{13}^-$    &   1.39E+04 &  3.94E+02 &   8.16  &   8 &   0.16\\
128  &$\Sigma$(1660)$P_{11}$      &   1.66E+04 &  1.20E+03 &   6.48  &   6 &   0.48\\
129  &$\Sigma$(1670)$D_{13}$      &   1.68E+04 &  6.00E+02 &   7.75  &   8 &   0.25\\
130  &$\Sigma$(1750)$S_{11}$      &   1.77E+04 &  1.10E+03 &   6.75  &   7 &   0.25\\
131  &$\Sigma$(1775)$D_{15}$      &   1.78E+04 &  1.20E+03 &   6.60  &   7 &   0.40\\
132  &$\Sigma$(1915)$F_{15}$      &   1.92E+04 &  1.20E+03 &   6.74  &   7 &   0.26\\
133  &$\Sigma$(1940)$D_{13}$      &   1.93E+04 &  2.25E+03 &   5.57  &   6 &   0.43\\
134  &$\Sigma$(2030)$F_{17}$      &   2.03E+04 &  1.75E+03 &   6.15  &   6 &   0.15\\
135  &$\Sigma$(2250)     &   2.25E+04 &  1.05E+03 &   7.27  &   7 &   0.27\\
136  &$\Xi^0$      &   1.31E+04 &  2.27E-11 &  54.81  &  55 &   0.19\\
137  &$\Xi^-$    &   1.32E+04 &  4.02E-11 &  53.97  &  54 &   0.03\\
138  &$\Xi$(1530)$P_{13}^0$    &   1.53E+04 &  9.10E+01 &  10.82  &  11 &   0.18\\
139  &$\Xi$(1530)$P_{13}^-$    &   1.54E+04 &  9.90E+01 &  10.69  &  11 &   0.31\\
140  &$\Xi$(1690)     &   1.69E+04 &  5.00E+02 &   8.09  &   8 &   0.09\\
\end{tabular}

\begin{tabular}{rlrrrrr}
Sl.  &   Particle   &    Mass    &  Width    &$n_{exact}$ & $n$ &  $D_{n1}$\\
No.  &Identification&    (MeV)   &  (MeV)    &         &     & ($n-n_{exact}$)\\
141  &$\Xi$(1820)$D_{13}$      &   1.82E+04 &  2.40E+02 &   9.49  &   9 &   0.49\\
142  &$\Xi$(1950)      &   1.95E+04 &  6.00E+02 &   8.02  &   8 &   0.02\\
143  &$\Xi$(2030)      &   2.03E+04 &  2.00E+02 &   9.97  &  10 &   0.03\\
144  &$\Omega^-$      &   1.67E+04 &  8.02E-11 &  53.30  &  53 &   0.30\\
145  &$\Omega$(2250)$^-$      &   2.25E+04 &  5.50E+02 &   8.42  &   8 &   0.42\\
146  &$\Lambda_{+c}$      &   2.29E+04 &  3.30E-08 &  44.81  &  45 &   0.19\\
147  &$\Lambda$(2625)$^+$      &   2.63E+04 &  3.20E+01 &  13.42  &  13 &   0.42\\
148  &$\Xi c^+$      &   2.47E+04 &  1.88E-08 &  45.76  &  46 &   0.24\\
149  &$\Xi c^0$      &   2.47E+04 &  5.20E+01 &  12.53  &  13 &   0.47\\
150  &$\Lambda^0 b$    &   5.64E+04 &  6.16E-09 &  48.66  &  49 &   0.34\\
151  &$\mu$      &   1.06E+03 &  3.00E-15 &  64.29  &  64 &   0.29\\
152  &$\tau$      &   1.78E+04 &  2.23E-08 &  45.02  &  45 &   0.02\\
153  &$W$      &   8.02E+05 &  2.08E+04 &   8.32  &   8 &   0.32\\
154  &$Z$      &   9.12E+05 &  2.49E+04 &   8.23  &   8 &   0.23\\
\end{tabular}

\newpage

\begin{center}
{\large{\bf TABLE 5}}
\end{center}

\begin{tabular}{rlrrrrr}
Sl.  &    ID    &    $E_r$    & $\Gamma$ & $\bigtriangleup \Gamma$& 
Log($E_r/\Gamma$) &  Dn  \\
No.  &          &          &  (Exp)   & (Errors) &          &      \\
     &          &          &          &          &          &      \\
  1  & Li-6     &    2.18  &   0.320  &   0.050  &    0.83  &    5 \\
  2  & Li-6     &    3.56  &   0 025  &   0.001  &    2.15  &   11 \\
  3  & Li-6     &     5.4  &   0.350  &   0.150  &    1.19  &    7 \\
  4  & Li-6     &    6.56  &   0.005  &          &    3.12  &   14 \\
  5  & Li-7     &  0.9779  &   0 268  &   0.030  &    0.25  &    1 \\
  6  & Li-7     &   4.629  &   1.000  &          &    0.67  &    4 \\
  7  & Li-7     &   7.475  &   0.006  &   0.000  &     3.1  &   14 \\
  8  & Li-7     &     8.9  &   0.089  &   0.007  &       2  &   10 \\
  9  & Li-7     &   11.13  &   0.093  &   0.008  &    2.08  &   10 \\
 10  & Li-8     &   2.258  &   2 000  &          &    0.05  &      \\
 11  & Li-8     &    3.21  &   4 000  &          &  -0. 10  &    1 \\
 12  & Li-8     &     6.4  &   1 000  &          &    0.81  &    5 \\
 13  & Li-8     &    6.51  &   0.040  &          &    2.21  &   11 \\
 14  & Li-8     &    6.53  &   0.310  &   0.005  &    1.32  &    7 \\
 15  & Be-7     &    4.55  &   0.298  &   0.025  &    1.18  &    7 \\
 16  & Be-7     &   7.185  &   1.800  &          &     0.6  &    4 \\
 17  & Be-7     &     9.9  &   1.200  &          &    0.92  &    6 \\
 18  & Be-7     &   10.79  &   0.836  &          &    1.11  &    6 \\
 19  & Be-7     &    11.4  &   0.100  &          &    2.06  &   10 \\
 20  & Be-8     &     2.9  &   7.000  &          &   -0.38  &    1 \\
 21  & Be-8     &    4.57  &   0.010  &   0.001  &    2.66  &   12 \\
 22  & Be-8     &   16.63  &   0.190  &          &    1.94  &   10 \\
 23  & Be-8     &   16.92  &   0.097  &   0.011  &    2.24  &   11 \\
 24  & Be-8     &    17.6  &   0.083  &   0.010  &    2.33  &   11 \\
 25  & Be-8     &   18.15  &   0.270  &   0.020  &    1.83  &    9 \\
 26  & Be-8     &   19.05  &   1.160  &          &    1.22  &    7 \\
 27  & Be-8     &   19.22  &   0.147  &          &    2.12  &   10 \\
 28  & Be-8     &   20.36  &   5.000  &          &    0.61  &    7 \\
 29  & Be-8     &    21.6  &   1.450  &   0.060  &    1.17  &    1 \\
 30  & Be-10    &   3.366  &   0.000  &   0.000  &    8.88  &   35 \\
 31  & Be-10    &   5.959  &   0.000  &          &    8.86  &   35 \\
 32  & Be-10    &   6.178  &   0.000  &          &    9.67  &   37 \\
 33  & Be-10    &   7.377  &   0.016  &          &    2.66  &   12 \\
 34  & Be-10    &   7.548  &   0.006  &          &     3.1  &   14 \\
 35  & Be-10    &    9.27  &   0.100  &          &    1.97  &   10 
\end{tabular}

\begin{tabular}{rlrrrrr}
Sl.  &    ID    &    $E_r$    & $\Gamma$ & $\bigtriangleup \Gamma$& 
Log($E_r/\Gamma$) &  Dn  \\
No.  &          &          &  (Exp)   & (Errors) &          &      \\
     &          &          &          &          &          &      \\
 36  & Be-10    &     9.4  &   0.400  &          &    1.37  &    7 \\
 37  & Be-10    &   17.79  &   0.110  &   0.035  &    2.21  &   11 \\
 38  & Be-10    &   18.47  &   0.500  &          &    1.57  &    8 \\
 39  & B-10     &  0.7173  &   0.000  &   0.000  &   12.04  &   46 \\
 40  & B-10     &    1.74  &   0.000  &   0.000  &     8.6  &   34 \\
 41  & B-10     &   2.154  &   0.000  &   0.000  &    9.65  &   37 \\
 42  & B-10     &   3.585  &   0.000  &   0.000  &    8.72  &   34 \\
 43  & B-10     &   4,774  &   0.001  &          &    3.68  &   16 \\
 44  & B-10     &   5.114  &   0.001  &          &    3.63  &   16 \\
 45  & B-10     &   5.166  &   0.000  &          &    6.24  &   25 \\
 46  & B-10     &   5.183  &   0.120  &   0.020  &    1.64  &    9 \\
 47  & B-10     &   5.923  &   0.005  &          &    3.07  &   14 \\
 48  & B-10     &   6.029  &   0.001  &          &    3.78  &   17 \\
 49  & B-10     &   6.133  &   0.005  &          &    3.09  &   14 \\
 50  & B-10     &   6.566  &   0.070  &          &    1.97  &   10 \\
 51  & B-10     &   6.884  &   0.140  &          &    1.69  &    9 \\
 52  & B-10     &       7  &   0.095  &   0.010  &    1.87  &    9 \\
 53  & B-10     &    7.43  &    0.14  &   0.030  &    1.72  &   10 \\
 54  & B-10     &   7.468  &    0.08  &          &    1.97  &   10 \\
 55  & B-10     &   7.479  &   0.072  &   0.004  &    2.02  &   10 \\
 56  & B-10     &   7.561  &   0.003  &   0.000  &     3.4  &   15 \\
 57  & B-10     &    7.62  &   0.225  &   0.050  &    1.53  &    8 \\
 58  & B-10     &    7.77  &    0.21  &   0.060  &    1.57  &    8 \\
 59  & B-10     &    8.07  &     0.8  &   0.200  &       1  &    6 \\
 60  & B-10     &   8.892  &   0.084  &   0.007  &    2.02  &   10 \\
 61  & B-10     &   8.896  &   0.036  &   0.002  &    2.39  &   11 \\
 62  & B-10     &     9.7  &     0.6  &          &    1.21  &    7 \\
 63  & B-10     &   10.83  &     0.5  &          &    1.34  &    7 \\
 64  & B-10     &    18.6  &     0.5  &          &    1.57  &    8 \\
 65  & B-10     &    18.8  &     0.6  &          &     1.5  &    8 \\
 66  & B-10     &    19.3  &    0.35  &          &    1.74  &    9 \\
 67  & C-10     &    10.2  &     1.5  &          &    0.83  &    5 \\
 68  & Al-25    &     2.5  &   0.001  &          &     3.4  &   15 \\
 69  & Al-25    &    2.69  &   0.000  &          &    3.84  &   17 \\
 70  & Al-25    &    2.74  &   0.001  &          &    3.32  &   15 
\end{tabular}

\begin{tabular}{rlrrrrr}
Sl.  &    ID    &    $E_r$    & $\Gamma$ & $\bigtriangleup \Gamma$& 
Log($E_r/\Gamma$) &  Dn  \\
No.  &          &          &  (Exp)   & (Errors) &          &      \\
     &          &          &          &          &          &      \\
 71  & Al-25    &    3.49  &   0.010  &          &    2.54  &   12 \\
 72  & Al-25    &    3.72  &   0.000  &          &    4.09  &   18 \\
 73  & Al-25    &    3.84  &   0.036  &          &    2.03  &   10 \\
 74  & Al-25    &    4.05  &    0.01  &          &    2.61  &   12 \\
 75  & Al-25    &    4.22  &       0  &          &    4.55  &   20 \\
 76  & Al-25    &    4.59  &       0  &          &    3.99  &   21 \\
 77  & Al-25    &     4.9  &    0.01  &          &    2.69  &   13 \\
 78  & Al-25    &    5.06  &   0.010  &          &     2.7  &   13 \\
 79  & Al-25    &     5.1  &    0.05  &          &    2.01  &   10 \\
 80  & Al-25    &     5.3  &     0.2  &          &    1.42  &    8 \\
 81  & Al-25    &    7.32  &     0.1  &   0.020  &    1.86  &    9 \\
 82  & Al-25    &    7.78  &    0.34  &   0.050  &    1.36  &    7 \\
 83  & P-29     &   4.342  &   0.053  &   0.003  &    1.91  &   10 \\
 84  & P-29     &   4.765  &   0.016  &   0.001  &    2.49  &   12 \\
 85  & P-29     &   4.968  &   0.004  &          &    3.09  &   14 \\
 86  & P-29     &    5.53  &   0.425  &   0.050  &    1.11  &    6 \\
 87  & P-29     &    5.74  &   0.013  &   0.001  &    2.66  &   12 \\
 88  & P-29     &   5.968  &    0.01  &   0.002  &     2.8  &   13 \\
 89  & P-29     &   6.195  &   0.095  &   0.006  &    1.81  &    9 \\
 90  & P-29     &   6.329  &   0.073  &   0.005  &    1.94  &   10 \\
 91  & P-29     &    6.59  &     0.2  &   0.020  &    1.52  &    8 \\
 92  & P-29     &   6.836  &   0.005  &   0.000  &    3.14  &   14 \\
 93  & P-29     &   6.956  &    0.12  &   0.010  &    1.76  &    9 \\
 94  & P-29     &   7.024  &     0.1  &   0.008  &    1.85  &    9 \\
 95  & P-29     &   7.963  &   0.008  &   0.001  &    2.95  &   14 \\
 96  & P-29     &   7.513  &   0.007  &   0.003  &    3.03  &   14 \\
 97  & P-29     &    7.62  &   0.165  &   0.025  &    1.66  &    9 \\
 98  & P-29     &    7.74  &   0.002  &          &    3.59  &   16 \\
 99  & P-29     &    7.92  &   0.014  &   0.004  &    2.75  &   13 \\
100  & P-29     &    7.97  &   0.125  &   0.025  &     1.8  &    9 \\
101  & P-29     &    8.08  &   0.036  &   0.010  &    2.35  &   11 
\end{tabular}

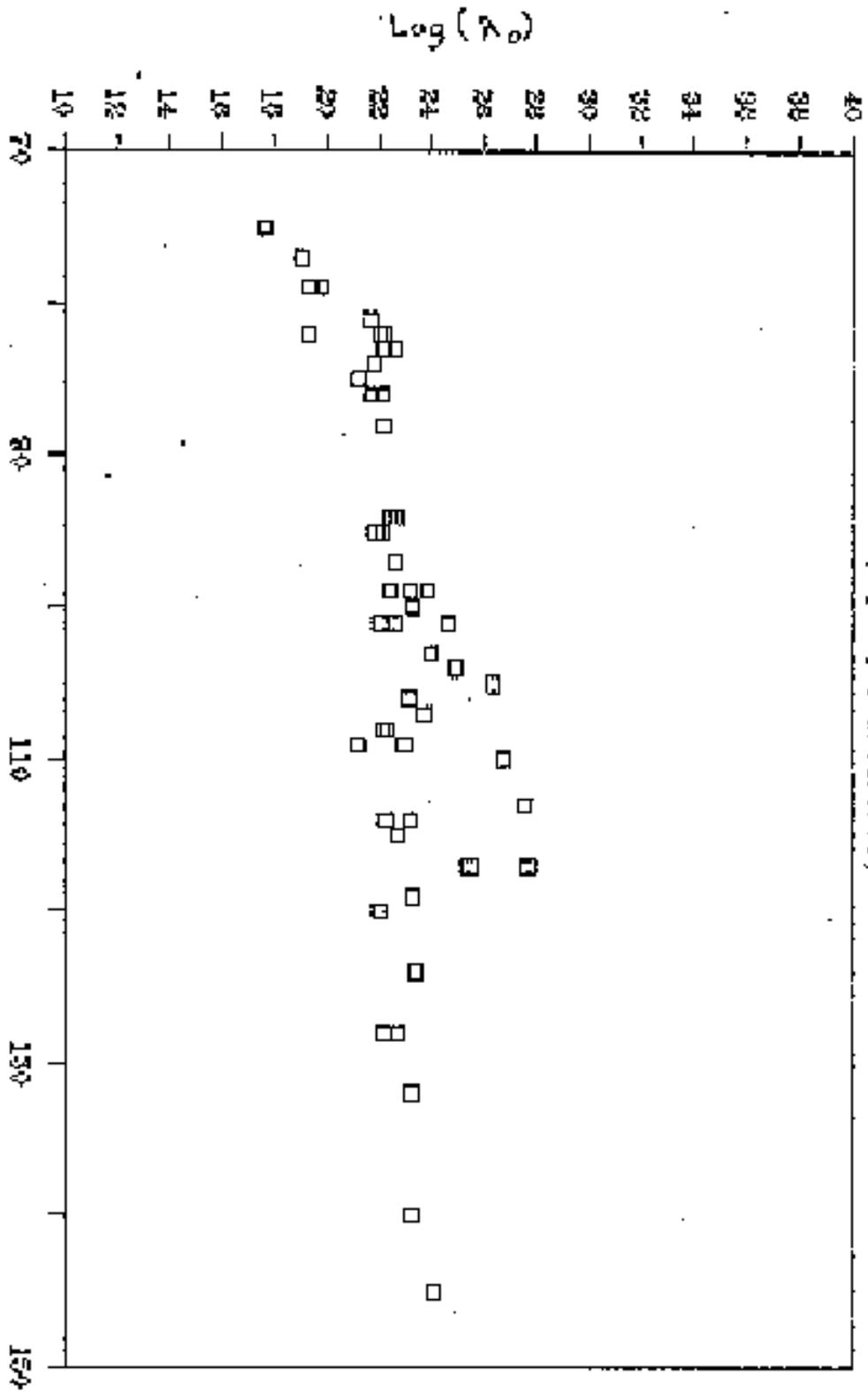
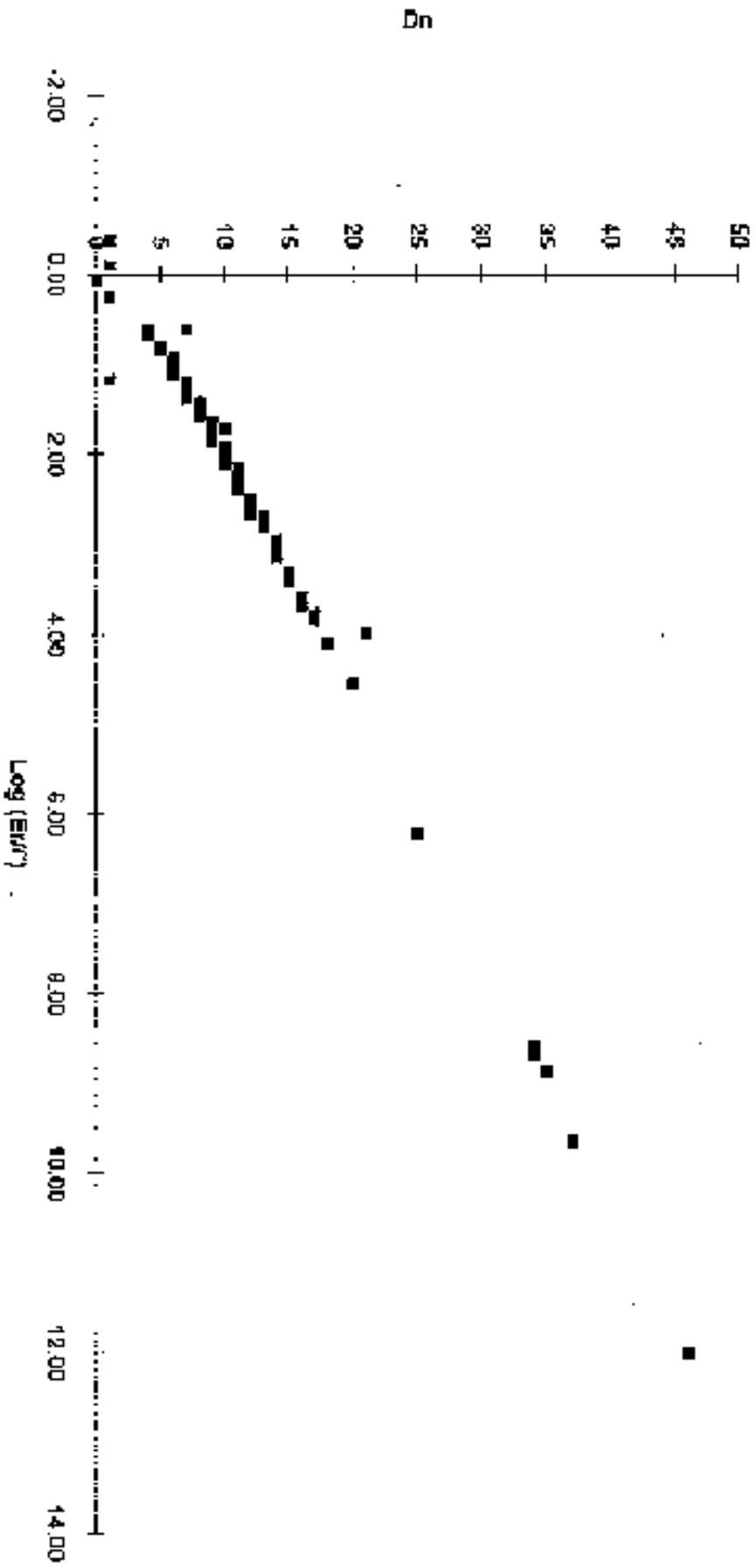
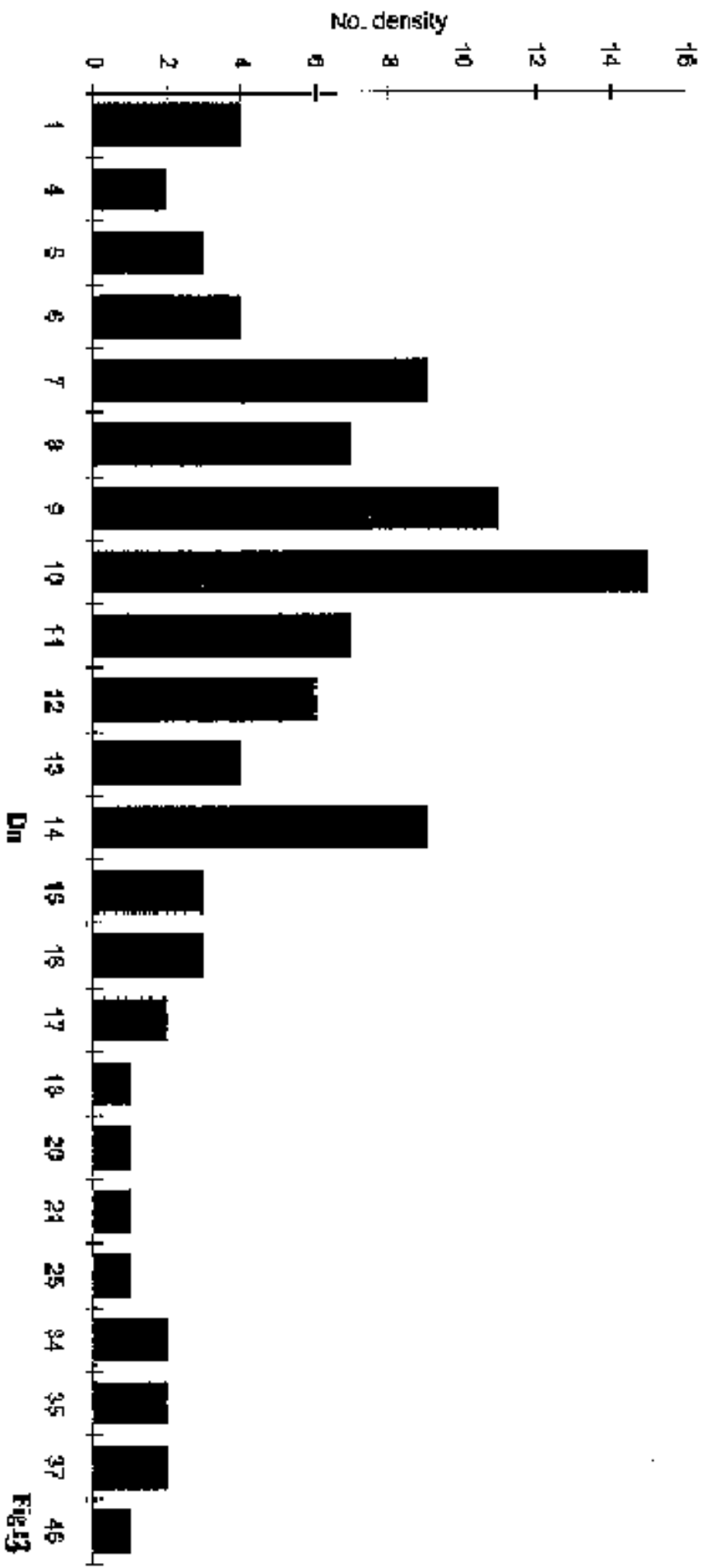
\end{document}